# Bloom-epistemic and sentiment analysis hierarchical classification in course discussion forums

Hapnes Toba[1], Yolanda Trixie Hernita[2], Mewati Ayub[1], Maresha Caroline Wijanto[2]
[1]Master of Computer Science Study Program, Faculty of Information Technology, Maranatha Christian University, Bandung, Indonesia
[2]Bachelor of Informatics Study Program, Faculty of Information Technology, Maranatha Christian University, Bandung, Indonesia

**ABSTRACT**

Online discussion forums are widely used for active textual interaction between lecturers and students, and to see how the students have progressed in a learning process. The objective of this study is to compare appropriate machine-learning models to assess sentiments and Bloom's epistemic taxonomy based on textual comments in educational discussion forums. The proposed method is called the hierarchical approach of Bloom-Epistemic and Sentiment Analysis (BE-Sent). The research methodology consists of three main steps. The first step is the data collection from the internal discussion forum and YouTube comments of a Web Programming channel. The next step is text preprocessing to annotate the text and clear unimportant words. Furthermore, with the text dataset that has been successfully cleaned, sentiment analysis and epistemic categorization will be done in each sentence of the text. Sentiment analysis is divided into three categories: positive, negative, and neutral. Bloom's epistemic is divided into six categories: remembering, understanding, applying, analyzing, evaluating, and creating. This research has succeeded in producing a course learning subsystem that assesses opinions based on text reviews of discussion forums according to the category of sentiment and epistemic analysis.

*Corresponding Author:*

Hapnes Toba
Computer Science Study Program, Faculty of Information Technology, Maranatha Christian University
Jalan Suria Sumantri No. 65, Bandung 40164, West Java, Indonesia
Email: hapnestoba@it.maranatha.edu

## 1. INTRODUCTION

Online discussion forums are one of the media that are currently being used by people to communicate with each other. Through interaction in discussion forums, it is very easy for discussion members to be able to develop a cooperative attitude and think critically in a forum [1]. A discussion forum is a positive tool for communicating with one another to share ideas and opinions. One example of the benefits of having a discussion forum for online learning is in the academic world. Learning activities can be more effective because students can solve problems through group discussions with lecturer observations during group discussions [2]. However, in the current e-learning environment, discussion forums have not been optimally used. Therefore, there is a need for collaboration between lecturers and students, as members of a social network, so that the discussion forum can run well [3]. In the discussion forums, lecturers can observe textual interactions between students and students can use the discussion forums in communicating with other students in the group. Students can express their opinions to each other, find solutions to problems, and develop each other's abilities, attitudes, and forms of positive behavior [4].

Discussion forums are part of the collaborative learning strategy. The purpose of such a collaborative learning model is to improve the ability of those who do not understand a study material perfectly. Students can share and interact with each other with different thoughts, opinions, and interpretations of learning materials and assignments. Discussion forum datasets are useful for analyzing each interaction between group members [5]. Given this situation, our study uses data from chat history in online learning environments.

One way to determine the attitude or nature of participants in the discussion through the history of textual interactions is to categorize the history of the conversation by sentiment [2], [6]. The purpose of sentiment analysis is to analyze the nature of each text message, the nature of sentiment analysis is primarily divided into three groups, namely negative, neutral, and positive. The benefit of this analysis is that it makes it possible to determine to what extent opinions are dispersed among the participants in the discussion to find solutions in a learning phase [7], [8]. Bloom's taxonomy may also serve as a standard to achieve the outcomes of the discussion forum, which is divided into six categories, namely remembering, understanding, applying, analyzing, evaluating, and creating [9], [10].

In learning environments, the automatic analysis of opinions [5]–[7] and Bloom's epistemic taxonomy [11]–[13] will be valuable for lecturers to adjust or extend the study materials based on students' comments. However, in recent years, the two approaches have evolved independently. To the extent of the knowledge, the research will fill the gap by combining epistemic analysis and sentiment analysis in one framework. The researchers will also demonstrate the features of machine learning algorithms that will be useful for each step of the analysis.

## 2. RESEARCH METHOD
### 2.1. Research framework

In general, the research framework involves four main phases. Figure 1 illustrates the research steps. The initial stage is the stage of collecting data from primary data sources, i.e., user comments in the Indonesian language from a programming course channel on YouTube. The data is gathered and will be processed during the annotation phase. The annotation step represents the data preprocessing and data labeling step. Data preprocessing is a process that must be performed as part of a data mining process so that the data to be used can be generated as necessary.

During the process of extracting sentences or words from the dataset, there are several stages of preprocessing, namely tokenization, filtering, and labeling [14]. After the preprocessing steps, the dataset will be annotated. Each text that appears in the forum will be annotated manually by three annotators (human assessment) depending on the sentiment class and epistemic category of the Bloom taxonomy [15], [16].

In the model creation phase, the concept of a two-step hierarchical classification is used to predict Bloom's epistemic and sentiment analysis (BE-Sent) on the forum datasets [17]. There are therefore two groups of classes: sentiment and epistemic. This research compares the random forest (RF) and long-short term memory (LSTM) methods that are used to learn model scenarios. During the evaluation phase of the model, the calculation of model performance is carried out by calculating the accuracy and confusion matrix analysis.

### 2.2. Contributions

The key contributions to this research include the following: i) We offer a new approach to analyzing the learning progress of students through interaction in discussion forums. To achieve this, we introduce the BE-Sent machine-learning model to predict Bloom's epistemic taxonomy and combine it with sentiment analysis to predict the students' opinions regarding the subject discussed in the forum; ii) We conducted a thorough analysis to identify challenging cases in the grouping of Bloom categories with the machine learning model mentioned; iii) We show how the BE-Sent model can be integrated into a course learning system (CLS).

### 2.3. Sentiment analysis

Sentiment analysis is also referred to as opinion mining [7]. It is a field of study that analyzes the opinions, judgments, and emotions of people towards different aspects such as products, services, organizations, individuals, problems, events, topics, and their related attributes [6]. Sentiment analysis has the objective to identify and evaluate a portion of text that expresses positive sentiments, negative sentiments, or neutral sentiments.

Positive sentiment is expressed when the text states happiness, approval, or agreement. For example, "I had a fantastic experience programming in Java language", would be a positive sentiment. Negative sentiment is expressed when the text states sadness, disapproval, or disagreement. For example, "I was extremely not satisfied with the service at the faculty", would be a negative sentiment. Neutral sentiment is expressed when the text states neither positive nor negative emotion. For example, "The source code is long", would be a neutral sentiment.

Sentiment analysis aims thus to find the value of emotional polarity in the text so that the polarity in each discussion conversation text can be assessed and classified. The generic approach used to analyze sentiment can be divided into three categories, namely machine learning [18], lexicon-based [19], and hybrid approach [20]. The categories are based on the nature of the model forming.

Sentiment analysis based on machine learning is a computational approach that determines the sentiment expressed in a piece of text by using machine learning algorithms. In this approach, a model is trained on a large dataset of labeled text samples (e.g., positive, negative, neutral) to learn patterns and relationships between the words, phrases, and sentiments. During the testing phase, the model is applied to classify new, unseen text data into one of the predefined sentiment categories. There are several machine learning algorithms used in sentiment analysis, such as naïve Bayes, support vector machines (SVM), and neural networks. In lexicon-based sentiment analysis, a list of words and phrases that have been annotated with sentiment polarity information (e.g., positive, negative, or neutral) is created, like a dictionary. To determine the sentiment of a piece of text, the words in the text are matched with the words in the sentiment lexicon. Finally, their sentiment polarity scores are aggregated to compute the overall sentiment of the text.

A hybrid approach to sentiment analysis aggregates multiple methods to improve the accuracy and robustness of sentiment analysis. For example, a hybrid approach can use a combination of lexicon-based sentiment analysis and machine learning-based sentiment analysis to take advantage of the strengths of both methods. The lexicon-based approach can provide fast and broad sentiment information, while the machine learning approach can provide more nuanced and context-aware sentiment information. The hybrid approach can achieve precision and robustness superior to one or the other method alone and is widely used in practical applications of sentiment analysis.

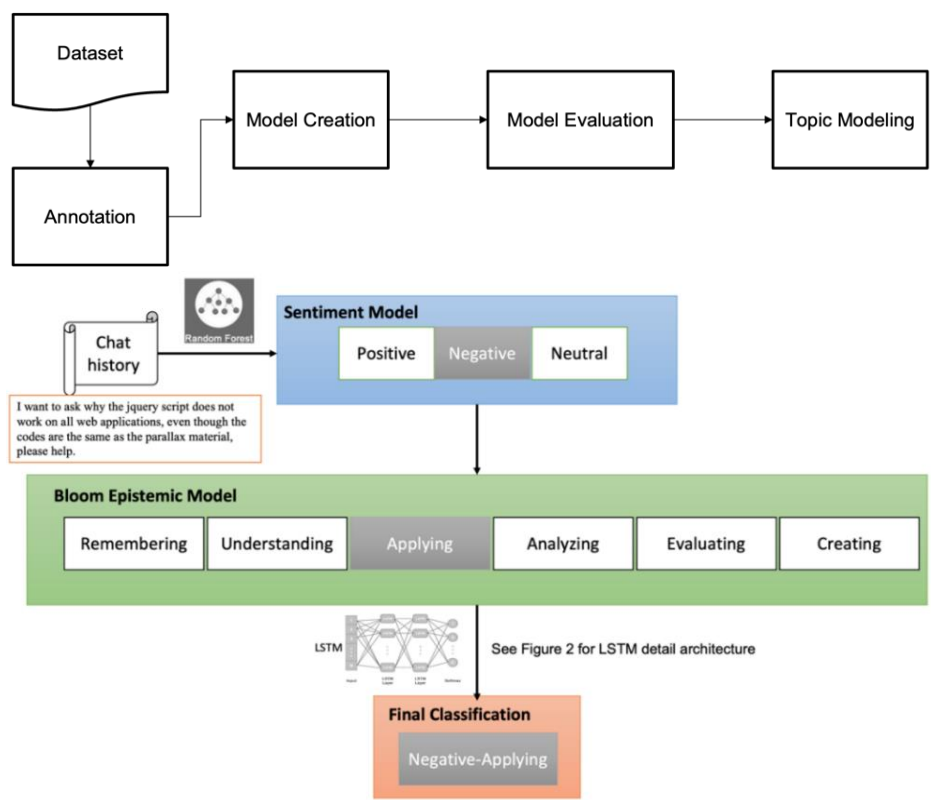

Figure 1. The general research framework and the concept of two-step hierarchical classification during model creation

**2.4. Bloom's taxonomy for epistemic analysis**

Bloom's taxonomy was developed in 1956 by Bloom [21]. Taxonomy is a system that underlies classification based on scientific research data. The epistemic cognitive domain of Bloom's taxonomy is an aspect of ability related to aspects of knowledge, reasoning, or thought which is divided into six categories, as given in Table 1 [22]. Bloom's taxonomy is identified with some influential verbs at each of its levels.

Table 1. The Bloom's taxonomy

| Taxonomy | Description | Examples of influential verbs |
|---|---|---|
| Remembering | The ability in the form of memory to remember (recall) or recognize (recognition) such as terminology, definitions, facts, ideas, methodologies, basic principles, and so on. | Define, identify, describe, recognize, explain |
| Understanding | The ability to understand and capture the meaning of what is learned. The existence of the ability to decipher, and change the data presented in a certain form to another form. | Summarize, interpret, classify, compare, extract |
| Applying | The ability to apply a method to handle a new case. Application of an idea, procedure, theory, and so on. | Solve, change, relate, use, discover |
| Analyzing | The ability to break complex information into small parts and be able to relate it to other information so that it can be understood better. | Illustrate, distill, conclude, categorize, connect |
| Evaluating | The ability to recognize data or information that must be obtained to produce a solution that is needed. | Criticize, reframe, appraise, value, grade |
| Creating | The ability to judge an argument regarding something that is understood, done, analyzed, and produced. | Design, modify, develop, collaborate, invent |

**2.5. Random forest classifier**

As an advanced machine-learning algorithm, RF classifiers can be used for sentiment analysis based on the decision tree algorithm [23]. It is also used to perform classification and regression. Formally, an RF is a combination of several 'good' decision tree models which are then unified as one big model. RF implements bootstrap sampling to build prediction trees. Each decision tree predicts with a random predictor and the accuracy value gets better if the number of trees used is large.

**2.6. Long short-term memory**

Long short-term memory (LSTM) is one of the powerful classification methods in deep learning. LSTM architecture has been developed as a solution to the problem of vanishing and exploding gradients encountered in recurring neural networks (RNNs) [24]. A vanishing gradient is caused due to the slowly decreased and unchanged weight values in the last layer thus causing it to never get a better or convergent result. On the other hand, too many increasing values of gradient cause the weight values in several layers to also increase, and thus the optimization algorithm becomes divergent which is called an exploding gradient.

LSTM has a chain-like structure where in each cell there are three gates, namely the forget gate, input gate, and output gate. LSTM has been proven to be effective for text mining solutions [25]. A simplified form of an LSTM architecture can be seen in Figure 2. A forget gate is a gate that manages the deletion of previously stored memory and at this gate will determine whether the information from input and output 1 will be allowed to continue the next process or not. This layer will produce output between 0 and 1. Output 0 is information that will be forgotten while output 1 is information that will not be forgotten and is allowed to pass.

The input gate determines which parts are to be updated and manages the storage of new information. The input gate has two parts, i.e., the neuron layer with the sigmoid activation function, and the neuron layer with the tanh activation function. The output gate decides what is to be produced as the final decision in each neuron. At the output gates, there are two gates to be implemented. Firstly, it will be decided which value in the memory cell will be supplied using the sigmoid activation function. Secondly, a value will be placed in the memory cell using the activation function tanh.

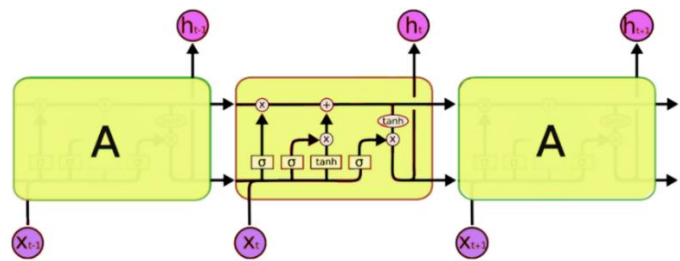

Figure 2. A generic layout of an LSTM cell architecture

**2.7. YouTube application programming interface**

YouTube application programming interface (API) is a public service provided by YouTube for programmers to create some interactions with video resources on YouTube channels [26]. Developers need to create some settings on Google Cloud Console to use the YouTube API [27]. An API is a very useful method for connecting applications or websites.

## 3. RESULTS AND DISCUSSION

The dataset is extracted from a publicly available discussion forum, i.e., the Pasundan University Web Programming Course on the YouTube Channel [28]. The extracted textual comments are taken from sessions 10 to 12. The general statistics for the dataset can be found in Table 2.

Table 2. General statistics of the dataset

| Number of videos | Number of main forums | Number of threads (Reply) | Number of chats | Number of words |
|---|---|---|---|---|
| 6 | 3,281 | 1,113 | 4,396 | 51,202 |

A discussion example, some replies, and the annotated BE-Sent are given in Table 2. The annotation is carried out by three students with a 0.72 inter-annotator agreement. For each chat, the annotator searches for the main word or phrase as an indicator to categorize it into sentiment (positive, negative, or neutral), and Bloom's categories: remembering, understanding, applying, analyzing, evaluating, and creating. For instance, in Table 3, the first chat contains the phrase '*tidak berfungsi*' (does not work), and the word '*coba*' (to try). Based on that, the chat is annotated with negative sentiment and Bloom's applying category.

Table 3. Forum example and the annotation

| Forum type | Textual content (in Bahasa Indonesia) | Textual content (translated into English) | Annotation |
|---|---|---|---|
| Main | *Ingin bertanya mengapa jquery tidak berfungsi di semua web padahal kode yang sudah coba dimasukkan sudah sama dengan materi parallax mohon bantuannya* | I want to ask why the jquery script does not work on all web applications, even though the codes are the same as the parallax material, please help. | Sentiment = negative<br>Epistemic = applying |
| Reply | *Cek urutan tag script jquery harus lebih dulu dari bootstrap* | Please check the order of the jquery script tags. Those must appear before the bootstrap. | Sentiment = neutral<br>Epistemic = analyzing |
| Reply | *Baik dicoba* | Noted, I will try the method. | Sentiment = positive<br>Epistemic = analyzing |
| Main | *Terima kasih tutornya* | Thank you for the tutorial. | Sentiment = positive<br>Epistemic = applying |
| Reply | *Sama-sama, semoga bermanfaat* | You're welcome, I hope it will be helpful | Sentiment = neutral<br>Epistemic = evaluating |

Further statistics of the training data are presented in Table 4. These statistics will be important to analyze whether or not we need to balance the distribution of classes. Based on the facts in Table 4, the synthetic minority oversampling technique (SMOTE) algorithm is applied to deal with imbalanced data during training. In total, our dataset consists of 4,396 chats to train and validate the models, and 100 chats to test models. During the training phase, we split the dataset into a hold-out composition of 70% train and 30% validation. We choose a 5-fold cross-validation (CV) setting as a baseline and compare the performance to the trained models. For the final evaluation, we collect a new dataset taken from the discussion forum in our internal CLS. There are 100 chats randomly chosen from the undergraduate 'Introduction of Web Programming' informatics subject during the even semester of the 2020/2021 academic year.

Table 4. Training data distribution

| Sentiment | | | Bloom | | | | | |
|---|---|---|---|---|---|---|---|---|
| Positive | Neutral | Negative | Remember | Understand | Apply | Analyze | Evaluate | Create |
| 1,742 | 2,332 | 322 | 36 | 2,688 | 1,599 | 24 | 24 | 25 |
| (39.63%) | (53.05%) | (7.32%) | (0.82%) | (61.15%) | (36.37%) | (0.55%) | (0.55%) | (0.57%) |
| | Total = 4,396 | | | | Total = 4,396 | | | |

### 3.1. Machine-learning performance

Several machine-learning experiments are carried out in this research: a 5-fold CV as the baseline, a hold-out of 70% training and 30% validation scenario, and a testing scenario from our internal discussion forum. The CV scenario is an acceptable method to tune and predict a global expectation of the model performance since there will be some training-testing iterations with randomly chosen instances from the dataset. The result of the CV experiment can be seen in Table 5.

Based on the baseline results of the 5 CV experiments in Table 5, the accuracy of the RF models is statistically significantly better than that of the LSTM models (p=0.05). Typically, an RF can trace the order of terms' occurrences. Thus, RF models will determine the order of important terms and assign them to the appropriate class. In addition, an LSTM model has the nature of randomness because of its artificial neural network characteristics. It is therefore not always capable of tracing the order of occurrences of terms.

Table 5. The baseline: 5-fold cross-validation accuracy performance

| Method | Sentiment | | Epistemic | |
|---|---|---|---|---|
| Iteration | RF (%) | LSTM (%) | RF (%) | LSTM (%) |
| 1 | 85.9 | 84.7 | 81.7 | 81.2 |
| 2 | 83.0 | 82.9 | 80.4 | 76.9 |
| 3 | 82.9 | 81.7 | 82.4 | 79.7 |
| 4 | 84.0 | 81.9 | 83.4 | 81.7 |
| 5 | 84.6 | 82.3 | 81.4 | 78.2 |
| Mean | 84.1 | 82.7 | 81.9 | 79.5 |
| Std. Dev. | 1.3 | 1.2 | 1.1 | 2.0 |

However, an LSTM model may be more flexible in general cases, as will be seen in the additional test dataset. An excerpt of an RF decision tree can be seen in Figure 3. In this example, we can track how the RF tree simulates the flow of a conversation. The branching of the tree is calculated by employing the Gini information gain [24], as shown in Table 6.

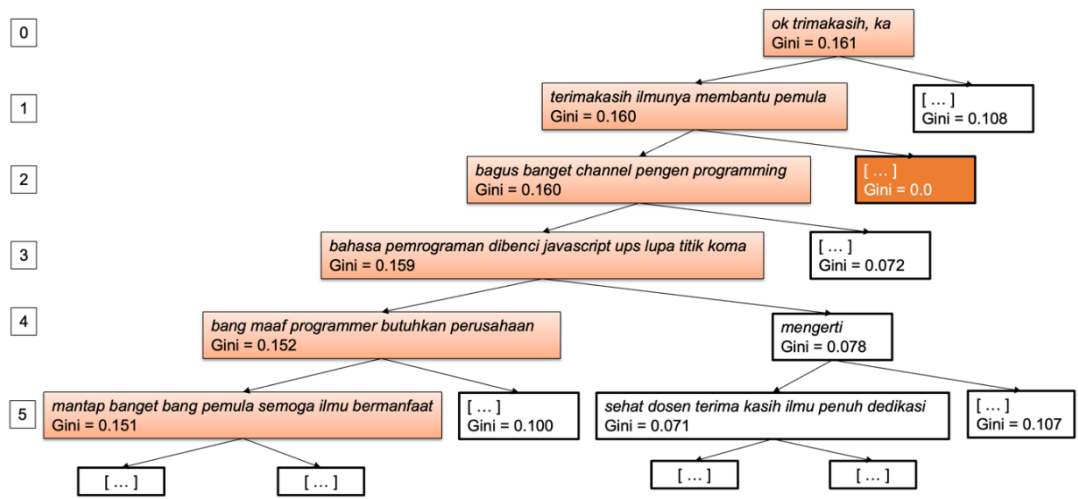

Figure 3. An RF Bloom decision tree excerpt for the 'understanding' classification

Table 6. An excerpt of the RF Bloom decision tree branch for the 'understanding' classification

| Level | Branch | Textual content (in Bahasa Indonesia) | Textual content (translated into English) | Gini |
|---|---|---|---|---|
| 0 | Root | *Ok trimakasih, ka* | Ok, thank you, sister. | 0.161 |
| 1 | Left | *Terimakasih ilmunya membantu pemula* | Thank you for the knowledge. It is very helpful for beginners. | 0.160 |
| 2 | Left | *Bagus banget channel pengen programming* | It is a very nice channel. Motivate me to learn to program. | 0.160 |
| 3 | Left | *Bahasa pemrograman dibenci javascript ups lupa titik koma* | I hate the javascript programming language. I always forget the semi-colon. | 0.159 |
| 4 | Left | *Bang maaf programmer butuhkan perusahaan* | Excuse me, brother. As a programmer, I need a work placement. | 0.152 |
| 5 | Left | *Mantap banget bang pemula semoga ilmu bermanfaat* | It is very cool, bro. Hopefully, the knowledge will be useful for beginners. | 0.151 |
| 4 | Right | *Mengerti* | I understand. | 0.078 |
| 5 | Right | *Sehat dosen terima kasih ilmu penuh dedikasi* | For all lecturers, thank you for your knowledge and dedication. Keep healthy. | 0.071 |

Further investigations during the experiments can be followed in Table 7. A hold-out composition of 70% training and 30% validation proportion is applied to the dataset. This table shows that each of the 'sentiment only' and 'epistemic only' models outperform the multilabel approach. This suggests that each classification problem has specific characteristics that need to be further classified with a dedicated model. Based on these results, we propose to use a hierarchical approach. First, we classified the sentiment, and based on the classified positive, negative, or neutral sentiment, a further epistemic classification process will be performed (Figure 1). The accuracy performance of the hierarchical classification method is comparable to the specific classification of sentiment and epistemic. This suggests that a hierarchical classification would be preferable in our case. The traversal from the root to the leaves can be followed in Table 6.

Table 7. Accuracy performance using 70%-30% hold-out validation scenarios (in %) of the proposed method

| Methods | 5-CV RF | RF | 5-CV LSTM | LSTM |
|---|---|---|---|---|
| Sentiment only | 84.1 +/-1.3 | 85.70 | 81.2 +/-1.2 | 82.30 |
| Epistemic only | 82.7 +/-1.1 | 82.10 | 76.4 +/-1.1 | 77.30 |
| Multilabel | 62.1 +/-1.1 | 63.20 | 54.3 +/-1.1 | 53.40 |
| Two-step hierarchical | 83.1 +/-1.2 | 84.35 | 78.2 +/-1.2 | 79.35 |

The next experiment involves testing the models with another equivalent dataset from our in-house CLS. The dataset consists of 100 randomly selected chats out of 14 regular course sessions. Comparable to the external dataset (YouTube), the chats within the internal dataset, are mostly manually classified as understanding (58%) and applying (42%). This implies that internal chats have also the tendency to correspond to Bloom's taxonomy for undergraduate studies. The objective of this experiment is to evaluate how convergent the general (i.e., publicly available) discussion forums are compared to the internal one. The result can be seen in Table 8.

From Table 8, it can be deduced that the performance of the test dataset decreased significantly. This is caused by the unseen terms in the internal dataset. Internal chats tend to have more formal greetings and questions regarding specific tasks or assignments. Another aspect is that internal chats are highly engaged to the students. While some students are more active than others, they tend to be more supportive and participatory.

Table 8. Accuracy performance of internal forum discussion test dataset

| Methods | RF (%) | LSTM (%) |
|---|---|---|
| Sentiment only | 67 | 53 |
| Epistemic only | 45 | 55 |
| Hierarchical | 62 | 49 |

The LSTM models are trained by using the default Keras library and hyper-parameters. We are optimizing some of the hyper-parameters with a random search strategy. Our focus is to find the optimal number of training epochs. During modeling, a sequential bidirectional LSTM approach is performed [18].

The first layer of the LSTM model is the word embedding layer which uses a 32-length vector composed by using the GloVe 500 words dimensionality reduction algorithm [5]. The next layer is the LSTM layer which has 100 neurons (each neuron is a single LSTM cell in Figure 2), with two hidden bidirectional layers. These layers serve for modeling memory cells. After that, the dense layer is constructed as an output layer with a sigmoid function. This sigmoid function is used to provide final rating labels. The LSTM architecture is shown in Figure 4.

The training-validation curves during the LSTM modeling phase are illustrated in Figures 5 and 6. Figure 5 shows the excerpt of LSTM's sentiment analysis during training validation and Figure 6 shows the epistemic performance of the LSTM. The number of optimum training epochs during sentiment analysis training is six, and seven for the epistemic analysis. The performance of the test-validation dataset is lower than the accuracy of the training. By further evaluation, the models seem to suffer from overfitting. This fact suggests that LSTM models are not general enough to catch variations of words occurring in chat histories.

Based on the results in Figures 5 and 6, exploring a deeper experiment to vary the length of the sentences in the forums would be important. This is important to anticipate the robustness of the time steps or the length of the sentiment sentences in our case, during the forming of LSTM models [25]. Another possibility to improve the LSTM models is to identify specific verbs according to Bloom's taxonomy from an extensive lexicon as guidance during classification in an encode-decoder scenario [29].

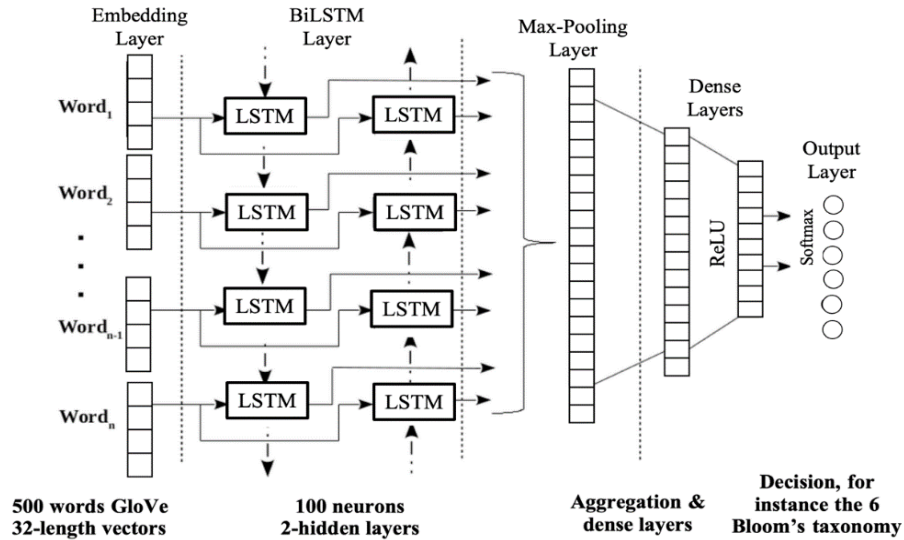

Figure 4. The overall architecture of the LSTM training model

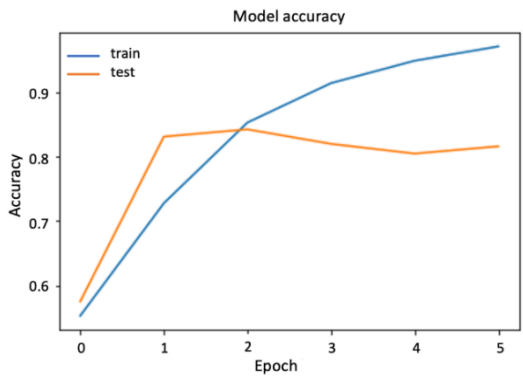

Figure 5. Sentiment analysis LSTM training-validation curve

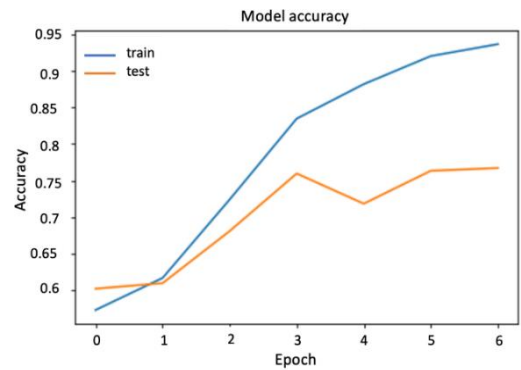

Figure 6. Epistemic analysis LSTM training-validation curve

### 3.2. Classification ambiguity

In this subsection, we further analyze the epistemic model performance by using the test dataset, which includes 100 chats from our internal CLS. The dataset is mainly annotated in the understanding (58%) and applying (42%) epistemic class. In our opinion, these two classes will also reflect the competencies of undergraduate students in general, and thus it is worth exploring how the RF and LSTM performed on the dataset. The accuracy performance of these two models can be seen in the confusion matrix in Table 9.

An interesting fact in Table 9 is that the RF model fits the 'understanding' class, while the LSTM model fits the 'applying' class. The false-positive rate of the RF model is statistically significantly higher than that of the LSTM model. In our observation, this implies that the LSTM has more ability to catch the randomness of word occurrences [24]. On the other hand, the RF model has more ability to predict the occurrences of some regularities in word occurrences [23]. This fact can be inferred from the first row in Table 7, which shows that more cases are classified as true. Some examples that show the cases of randomness in epistemic analysis and the regularity of sentiment analysis can be seen in Table 10.

Table 9. Confusion matrix of RF and LSTM accuracy in the understanding and applying classes (%)

| RF | Understanding | Applying | LSTM | Understanding | Applying |
|---|---|---|---|---|---|
| Understanding | 36 | 22 | Understanding | 32 | 26 |
| Applying | 33 | 9 | Applying | 19 | 23 |

Table 10 shows the examples of specific words that indicate their occurrences in specific sentiment analysis classes. On the other hand, the same words could be classified into different classes of epistemic analysis. Based on this observation, the LSTM model might be enhanced by first identifying important verbs to indicate a specific level in Bloom's taxonomy [29]. An example of model integration in a real-case scenario is depicted in Figure 7. The sentences in the yellow boxes of Figure 7 are the English translation of the discussion interactions. The predicted sentiments' labels are as: *netral* → neutral; *negatif* → negative; *positif* → positive. The predicted Bloom's labels are as: app → applying; rem → remembering; und → understanding.

Table 10. Examples of word randomness and regularity occurrences in epistemic and sentiment analysis

| Type of word occurrences | Specific Word example (Indonesian) | Specific Word example (English translation) | Sentiment class | Epistemic class |
| --- | --- | --- | --- | --- |
| Formal greetings | *Selamat pagi/siang/sore* | Good morning/afternoon/evening. | Positive | Understanding, and applying |
| | *Terima kasih* | Thank you. | | |
| Negation | *Bukan hal itu* | Not that kind of thing. | Negative | Understanding |
| | *Tidak bisa* | It cannot be done. | | |
| Technological terms | *Unduh* | To download. | Neutral | Understanding, and applying |
| | *Alat edit teks* | Text editor. | | |

Figure 7. Subsystem response according to the chat history in a discussion forum

### 3.3. Subsystem integration

Based on the RF and LSTM models that have been deployed, a new subsystem is constructed to improve an engagement-based learning management system (LMS) [30]. We integrated the RF model to predict the sentiment analysis, and the LSTM model to predict the epistemic. Following this enhancement, the lecturers can analyze the discussion forum in a deeper sense. With this functionality, they might estimate how far students have progressed within a session. Figure 7 shows an example of how the subsystem responded in a discussion forum. In the near future, we also plan to improve the LSTM model for epistemic analysis by incorporating word class identification, *i.e.*, verb classification in Bloom's taxonomy.

### 4. CONCLUSION

In this paper, the researchers have demonstrated the process of developing machine-learning models for conducting epistemic and sentiment analysis. There were two machine learning algorithms applied (RF and LSTM). In aggregating the two algorithms, a two-stage classification concept is used. The first stage is on the sentiment aspect and the next is on the epistemic aspect. The observations showed that the RF model tends to be more appropriate for analyzing sentiments because it can capture the occurrence of regular words. In contrast, the LSTM model is more likely to be adapted to predict epistemic classes with occurrences of words that tend to be more randomized. The model training results have successfully been implemented in an engagement-based LMS to help lecturers observe the progress of student learning through conversational history in discussion forums.

**ACKNOWLEDGEMENTS**

The research presented in this paper was supported by the Research Institute and Community Service (LPPM) at Maranatha Christian University, Bandung, Indonesia.

## BIOGRAPHIES OF AUTHORS

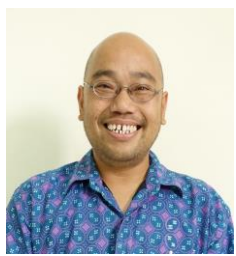

**Hapnes Toba** graduated in 2002 with a Master of Science from the Delft University of Technology in the Netherlands and completed his doctoral degree in computer science at Universitas Indonesia in 2015. He is an associate professor in the area of artificial intelligence and is interested in information retrieval, natural language processing, educational data mining, and computer vision. He has been a faculty member in the Faculty of Information Technology at Maranatha Christian University since 2003. He is also an active board member of the Indonesian Computational Language Association (INACL) and serves as the chair of the Information and Communication Technology Forum of the Association of Christian Universities and Colleges in Indonesia. He can be contacted by email at: hapnestoba@it.maranatha.edu.

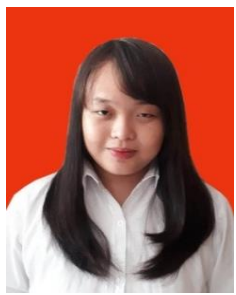

**Yolanda Trixie Hernita** is an alumnus of Maranatha Christian University. She is majoring in Informatics Engineering very interested in web programming and data analytics. Apart from the academic field, she also joined the Voice of Maranatha (VOM) organization on campus, from 2018 to 2019. At the recent moment, she serves as a data analyst at a national banking organization. She can be contacted by email at: 1872045@maranatha.ac.id.

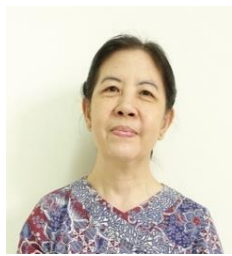

**Mewati Ayub** graduated with a Bachelor of Informatics from Bandung Institute of Technology (ITB) in 1986 and completed her master's degree at Bandung Institute of Technology in 1996, and her doctoral degree at Bandung Institute of Technology in 2006. She has been working as a faculty member in the Faculty of Information Technology at Maranatha Christian University since 2006. Her specialty is in the field of educational technology, software engineering, and data mining. She can be contacted by email at: mewati.ayub@it.maranatha.edu.

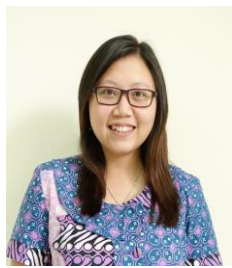

**Maresha Caroline Wijanto** is an alumnus of the Faculty of Information Technology, Maranatha Christian University, and graduated from the Bandung Institute of Technology (ITB) with her Master's Degree in Computer Science. She joined the Faculty of Information Technology at Maranatha Christian University in 2010. Her specialty is in the field of Natural Language Processing, Machine Learning, and Data Mining. She can be contacted by email at: maresha.cw@it.maranatha.edu.